\documentclass[twocolumn, twocolappendix]{aastex631}
\usepackage{amsmath,bm}
\usepackage{multirow}
\usepackage{enumitem}
\usepackage{color}
\usepackage{ulem}
\usepackage{CJKfntef,CJK}
\usepackage{makecell,ulem}

\newcommand\hh{$\rm H_2$}
\newcommand\coa{$\rm ^{12}CO$}
\newcommand\cob{$\rm ^{13}CO$}

\newcommand\kms{$\rm km\ s^{-1}$}
\newcommand\um{$\mu$m}
\newcommand\water{$\rm H_2O$}
\newcommand\chh{$\rm CH_2$}
\newcommand\chhh{$\rm CH_3$}

\begin{document}
\begin{CJK*}{UTF8}{gbsn}
\title{Dust Absorption towards Supernova Remnant W44}

\author[0000-0002-9776-5610]{Tian-Yu Tu (涂天宇)}
\affiliation{School of Astronomy \& Space Science, Nanjing University, 163 Xianlin Avenue, Nanjing 210023, China}
\email{tianyu.tu1105@outlook.com}

\author[0000-0001-9344-0096]{Adwin Boogert}
\affiliation{Institute for Astronomy, University of Hawaii, 2680 Woodlawn Drive, Honolulu, HI 96822, USA}

\author[0000-0002-4753-2798]{Yang Chen (陈阳)}
\affiliation{School of Astronomy \& Space Science, Nanjing University, 163 Xianlin Avenue, Nanjing 210023, China}
\affiliation{Key Laboratory of Modern Astronomy and Astrophysics, Nanjing University, Ministry of Education, Nanjing 210023, China}
\email{ygchen@nju.edu.cn}

\author[0009-0002-4427-6976]{Wenlang He (何文朗)}
\affiliation{School of Astronomy \& Space Science, Nanjing University, 163 Xianlin Avenue, Nanjing 210023, China}

\begin{abstract}
Supernova remnants (SNRs) can strongly affect the chemical composition of the interstellar dust.  
In this paper we investigate to what degree the dust and ices are modified by observing four stars expected to be absorbed by a giant molecular cloud interacting with SNR W44, using medium-resolution spectroscopy in 2--5 \um.  
Absorption from \water\ ice around 3.0 \um\ and aliphatic hydrocarbon dust around 3.4 \um\ were detected towards two stars, while probable CO ice at 4.67 \um\ towards one of them. 
Millimeter gas-phase CO $J=1$--0 lines and three-dimensional dust extinction maps show that the dense molecular gas associated with W44 dominates ($\gtrsim 60\%$) the total interstellar extinction ($A_K\sim 2.6$) along these two sightlines.  
The \water\ ice column densities are a factor of 1.5--3 lower than nearby MCs at similar extinctions, possibly because of the destruction of ice by shocks and cosmic rays (CRs) from W44, consistent with the low CO ice abundance relative to \water\ ($\lesssim 12\%$).  
One of the sightlines shows an unusually strong 3.4 \um\ aliphatic hydrocarbon absorption.  
If the carriers are located in diffuse dust along the sightline, unrelated to W44, its strength is $\sim4$ times larger than those typically observed for diffuse dust clouds. 
Alternatively, the carriers may be enhanced in the W44 environment.  
We discuss several possible explanations, including shock formation of aliphatic hydrocarbons in diffuse clouds associated with W44, contribution from aliphatic hydrocarbons in shocked and CR-bombarded molecular clouds, and changes in the extinction law due to the SNR interaction.

\end{abstract}


\section{Introduction} \label{sec:intro} 
Supernova remnants (SNRs) exert strong feedback to the interstellar medium (ISM) and serve as important regulators of galaxy evolution and star formation \citep{Pillepich_Simulating_2018}.  
When the blast waves of SNRs propagate into the interstellar molecular clouds (MCs), they can significantly alter the physical and chemical properties of the gas and dust inside the MCs.  
The physical and chemical effects of SNR shocks on molecular gas have been widely studied \citep[e.g.,][]{vanDishoeck_Submillimeter_1993,Reach_Excitation_1999,Brogan_OH_2013,Mazumdar_Submillimeter_2022a}. 
However, the change in the properties, especially the composition, of the dust in the MCs affected by SNRs are seldom studied.

\par

Dust grains in MCs consist of a core with silicates, carbon solids, hydrocarbons, metallic compound, etc. \citep[e.g.,][]{Draine_Physics_2011}, covered with an ice mantle of $\rm H_2O$, CO, $\rm CO_2$, $\rm CH_3OH$, etc. at different extinctions \citep[e.g.,][]{Boogert_Observations_2015}.  
The chemical composition of ice has been studied towards many quiescent MCs \citep[e.g.,][]{Boogert_Ice_2011,Chiar_Ices_2011,Whittet_Ice_2013,Boogert_Infrared_2013,Goto_first_2018,Chu_Observations_2020,Goto_Water_2021,Madden_Infrared_2022}, and recent observations by the James Webb Space Telescope revealed a wealth of complex molecular species on the ice mantle of dust grains in MCs \citep[e.g.,][]{McClure_Ice_2023}.  
The chemistry on dust grains is of great importance because it is crucial to the explanation of observed gas-phase molecular abundances \citep[e.g.,][]{Garrod_Formation_2006,Codella_Astrochemistry_2015} and this is the material that builds envelopes and protoplanetary disks of young stellar objects \citep{Caselli_Our_2012}.

\par

Shocks driven by SNRs can destruct the dust grain in the diffuse interstellar medium \citep[e.g.,][]{Barlow_destruction_1978,Jones_Grain_1996,Micelotta_Polycyclic_2010,Hu_Thermal_2019}. 
In dense MCs, the shocks are largely decelerated compared to those in diffuse gas, but can still destruct the dust grains through various processes such as sputtering, grain-grain collision, and vaporization
\citep[e.g.,][]{Caselli_Grain-grain_1997,Bergin_Postshock_1998,Jimenez-Serra_Parametrization_2008,Guillet_Shocks_2009,Guillet_Shocks_2011,Burkhardt_Modeling_2019}.
The silicon locked in the dust cores can be transformed to SiO molecules in the gas phase such as those which have been detected towards several SNRs and can serve as solid evidence for the presence of shock-MC interaction \citep[e.g.,][]{vanDishoeck_Submillimeter_1993,Nicholas_7_2012,Dumas_Localized_2014,Cosentino_Interstellar_2019,Cosentino_Negative_2022,Mazumdar_Submillimeter_2022a}.
The icy molecules adsorbed on the dust mantle can be released to the gas phase, either observed directly or reacting with other species in the hot shocked gas, leading to unusual molecular chemistry in the shocked MCs of SNRs \citep[e.g.,][]{Maxted_Ammonia_2016,Mazumdar_Submillimeter_2022a}.
Simulations have also found that although shocks can strongly destruct the dust mantle, ice can be re-formed efficiently on the grain surface in the cooling postshock region with different abundances from the preshock MCs \citep[e.g.,][]{Bergin_Formation_1999}.  
This postshock ice composition can be retained and inherited by the next-generation star formation \citep{Burkhardt_Modeling_2019}.

\par

Cosmic rays (CRs), of which the SNRs are important accelerators \citep[e.g.,][]{Aharonian_Gamma_2013}, can also affect the chemical properties of the dust.  
CR radiation can also induce the destruction of dust cores \citep{Barlow_destruction_1978}.  
The influence of CRs on dust mantles is complicated.  
They can dissociate the icy molecules on mantle along their trajectories, forming radicals and excited species which fuels the formation of more complex molecules without barriers \citep{Shingledecker_general_2018,Shingledecker_Cosmic-Ray-driven_2018,Arumainayagam_Extraterrestrial_2019}.
They can also release the icy molecules to the gas phase through various processes with various efficiency for different species \citep[e.g.,][]{Ivlev_Impulsive_2015,Dartois_Cosmic_2018,Wakelam_Efficiency_2021}.
These processes are suggested to be responsible for the gas-phase complex organic molecules detected in quiescent MCs \citep[e.g.,][]{Bacmann_Detection_2012}.

\par

W44 (G34.7$-$0.7) is a middle-aged SNR \citep[$\sim 7$--27 kyr, e.g.,][]{Wolszczan_Discovery_1991,Rho_X-Ray_1994,Park_Shells_2013} located at a distance of $\sim 3$ kpc \citep[e.g.,][]{Park_Shells_2013,Lee_High-resolution_2020}.  
It is believed to be interacting with its surrounding MCs, as evidenced by the 1720 MHz OH masers \citep{Claussen_Polarization_1997,Hoffman_OH_2005a}, class I $\rm CH_3OH$ maser \citep{McEwen_NH333_2016}, CO observations (including broadened line profiles with $\rm FWHM\gtrsim10$ \kms, dynamic analysis, enhanced high-to-low excitation line ratios, high-$J$ transitions, etc. \citep[e.g.,][]{Seta_Enhanced_1998,Seta_Detection_2004,Reach_Shocked_2005,Yoshiike_Neutral_2013,Sashida_Kinematics_2013,Anderl_APEX_2014}, SiO emission \citep[][]{Cosentino_Widespread_2018,Cosentino_Interstellar_2019}, infrared atomic, ionic and \hh\ lines \citep[e.g.,][]{Reach_Shockingly_1996,Reach_Infrared_2000,Lee_Near-infrared_2019,Lee_High-resolution_2020}, etc.  
It is also an accelerator of CRs with $\gamma$-ray emission \citep[][]{Ackermann_Detection_2013,Cardillo_supernova_2014,Peron_Gamma-Ray_2020}, enhanced CR ionization rate \citep[][]{Cosentino_Interstellar_2019}, and X-ray 6.4 keV Fe I K$\alpha$ line \citep[][]{Nobukawa_Evidence_2018}.  
Therefore, W44 is an ideal target for us to study the feedback of SNRs, especially on dust.

\par

Infrared spectroscopy of solid-state absorption features provides a direct and powerful technique for investigating the composition and evolution of dust in the interstellar medium (see \citet{Whittet_Dust_2022} for a comprehensive review).  
In this paper, we present new observations of dust absorption against four stars towards SNR W44.  
In Section \ref{sec:obs}, we describe the choice of the target sources and details about the observation and data reduction.  
The procedure of continuum determination and estimation of the optical depths and column densities of the detected dust and ice absorption features are described in Section \ref{sec:res}.  
In Section \ref{sec:disc}, we discuss the possible interpretation of the observational results of the dust and ice under the influence of W44.
Our conclusions are summarized in Section \ref{sec:con}.

\begin{deluxetable*}{ccccccccccc}[!t]
\tablecaption{Summary of the target stars 
\label{tab:info}}
\tablehead{
\colhead{ID} & \colhead{2MASS ID} & \colhead{\makecell{WISE1\\(mag)}} & \colhead{\makecell{WISE2\\(mag)}} & \colhead{\makecell{$J-K_s$\\(mag)}} & \colhead{Spectral Type} & \colhead{\makecell{$A_K$\\(mag)}} & \colhead{$\tau_{3.0}$} & \colhead{\makecell{$N(\rm H_2O_{ice})$\\($\rm 10^{17}\ cm^{-2}$)}} & \colhead{$\tau_{4.67}$$^c$} & \colhead{\makecell{$N(\rm CO_{ice})$$^c$\\($\rm 10^{16}\ cm^{-2}$)}}
}
\startdata
1$^a$ & 18563600+0119174 & 5.564 & 5.445 & 5.250 & M6.5S--M7SIII & 2.15 & $\leq0.038$ & $\leq0.63$ & --- & --- \\
2     & 18563962+0125425 & 6.601 & 6.132 & 5.097 & M6.5S--M7SIII & 2.57 (0.12) & 0.39 (0.03) & 6.48 (1.20) & $\le 0.20$ & $\le 6.91$\\
3$^b$ & 18565889+0127047 & 6.782 & 6.189 & 4.296 & M6e--M9eIII   & 1.61 (0.09) & $\leq0.10$ & $\leq1.7$ & --- & --- \\
4     & 18565584+0123043 & 6.751 & 6.482 & 5.181 & K4Ib-II       & 2.59 (0.13) & 0.27 (0.03) & 4.52 (0.84) & $\le 0.16$ & $\le 5.55$ \\ 
\enddata
\tablecomments{
Values given in the parentheses show the $1\sigma$ uncertainty. 
$a$: The continuum determination of star 1 is not satisfactory (see Section \ref{sec:conti}), so the best-fit spectral type and extinction may be problematic. The upper of CO ice is not calculated. 
$b$: No \water\ ice is detected towards star 3, so only an upper limit is provided. The optical depth of the CO ice cannot be determined because the observed flux is lower than the modeled flux (see Figure \ref{fig:star-specs}). 
$c$: The CO ice absorption is probably overlapped with the CO gas absorption, so the optical depth and column density should be regarded as upper limits (see Section \ref{sec:column}). 
}
\end{deluxetable*}

\section{Observations and data reduction} \label{sec:obs}
\begin{figure}[t]
    \centering
    \includegraphics[width=0.48\textwidth]{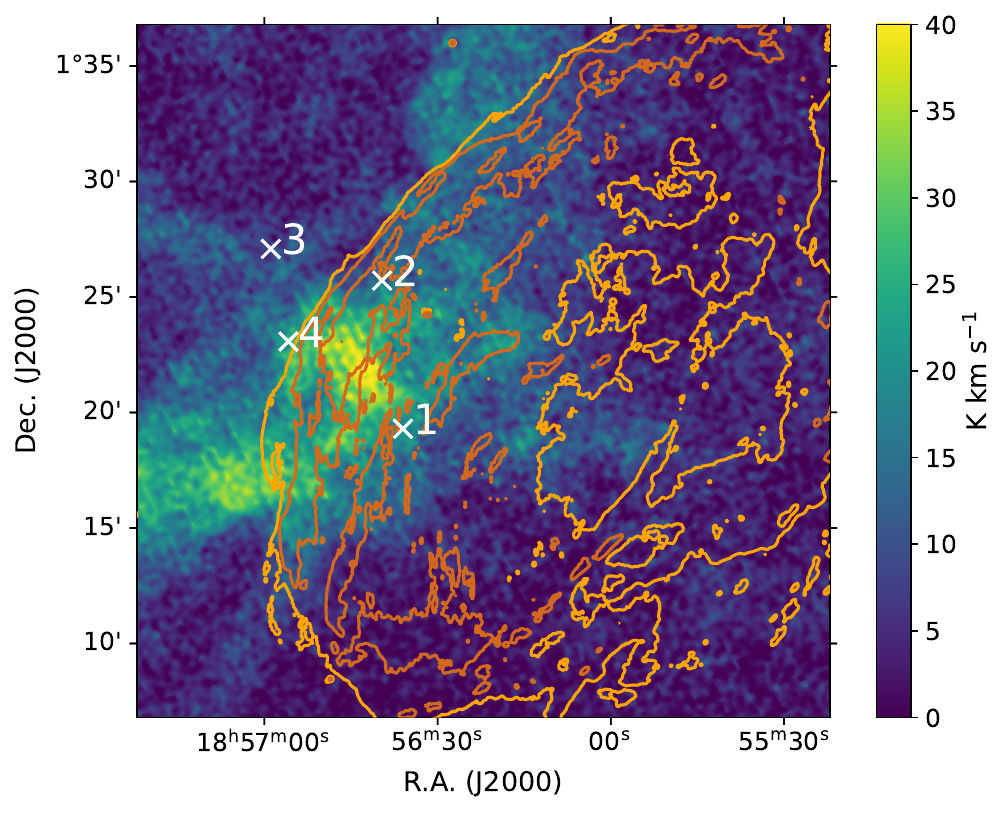}
    \caption{
    Integrated intensity map of FUGIN \cob\ 1--0 line in $+39$--$+47$ \kms\ covering the majority of the molecular emission associated with W44 \citep{Cosentino_Interstellar_2019}. 
    The white crosses mark the positions of the target stars, while the orange and dark orange contours show the SMGPS 1.3 GHz radio continuum of W44 in 2 and 8 mJy $\rm beam^{-1}$, respectively. 
    }
    \label{fig:13COmap}
\end{figure}

A careful selection of target stars is necessary to increase the probability to detect the dust and ice absorption in MCs associated with W44. 
The target are selected with the following criteria: 
(1) bright in the WISE2 \citep{Wright_Wide-field_2010} photometry (of magnitude $W2<6.5$ mag at 4.6 \um), 
(2) red in the 2MASS \citep{Skrutskie_Two_2006} photometry (with $J-K_s>4$ mag) to bias towards the heavily absorbed stars, and
(3) covered by the extended \cob\ 1--0 emission that is known to be associated with the SNR. 
Finally we chose four stars based on these criteria, and their information is listed in Table \ref{tab:info}. 
The positions of these stars, and their comparison with \cob\ 1--0 line emission and the radio boundary of W44, are shown in Figure \ref{fig:13COmap}. 

The observations were carried out with SpeX \citep{Rayner_SpeX_2003}, which is a 0.7--5.3 \um\ medium-resolution spectrograph mounted on the NASA Infrared Telescope Facility (IRTF) on Maunakea.  
The LXD\_long observation mode of SpeX results in a wavelength coverage of 1.98--5.3 \um, allowing us to simultaneously search for the stretching modes of water ice (at $\sim 3.0$ \um) and CO ice (at $\sim 4.67$ \um) as well as the absorption features of other dust species.  
A slit width of $0^{\prime\prime}.8$ is chosen to reach a spectral resolution of $R=\lambda/\Delta\lambda \sim 940$.  
The spectra were recorded by nodding the telescope along the slit every exposure to remove the sky
background emission.  
The standard star, $\lambda$ Aql with a B9V spectral type, was observed regularly at similar airmass to the observed targets for the telluric correction in the data reduction process.  
The observations were conducted on two half-nights in August 2024.

\par

Following the standard procedure, we reduced the raw data with Spextool v4.1, which is an IDL-based data reduction package \citep{Cushing_Spextool_2004}.  
The entire procedure includes flat field calibration, wavelength calibration, subtraction of sky background, extraction of the spectra, data cleaning and coadding, and telluric correction.  
The typical signal-to-noise ratios (S/N) of the reduced spectra vary from $\sim 35$ to $\sim 50$ in $L$ band (around $\sim 3.7$ \um) and from $\sim10$ to $\sim20$ in $M$ band (around $\sim 4.7$ \um). The reduced spectra are finally flux-calibrated with their 2MASS $K_s$ photometry.

\par

Supplementary data of the \coa\ 1--0 and \cob\ 1--0 line emission was retrieved from the FOREST unbiased Galactic plane imaging survey with the Nobeyama 45 m telescope \citep[FUGIN,][]{Umemoto_FOREST_2017}. 
The 1.3 GHz radio continuum image of W44 was taken from the SARAO MeerKAT 1.3 GHz Galactic Plane Survey \citep[SMGPS,][]{Goedhart_SARAO_2024}, and additional 1.4 GHz H\,I spectral line data was obtained from the Galactic Arecibo L-Band Feed Array H\,I survey \citep[GALFA-H\,I,][]{Peek_GALFAHI_2018}.
We used \textit{Python} package \textit{Astropy} for further analysis of the data. 

\section{Results} \label{sec:res}

\begin{figure*}[t]
    \centering
    \includegraphics[width=0.99\textwidth]{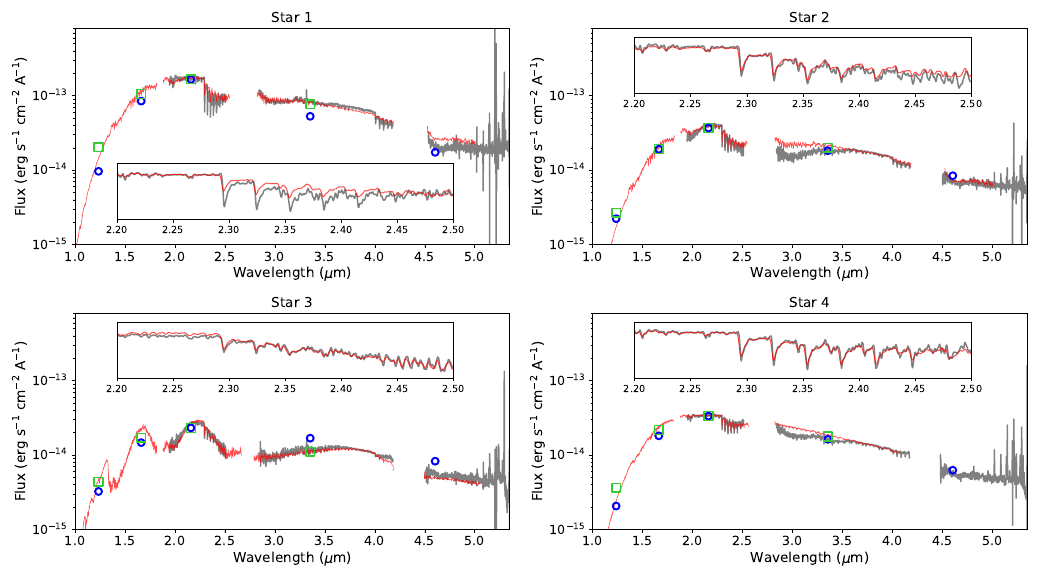}
    \caption{
    Observed spectra of the target stars (grey lines), as well as WISE and 2MASS photometry (open blue circles, $J$, $H$, $K_s$, $W1$, and $W2$ from short to long wavelengths). 
    The best-fit reddened photospheric model spectra are presented in thin red lines, and the modeled $J$, $H$, $K_s$, and $W1$ photometry is shown in open green boxes.
    A zoom-in plot showing the CO first overtone is inserted in each panel. 
    }
    \label{fig:star-specs}
\end{figure*}

\subsection{Continuum Determination} \label{sec:conti} 

The observed spectra are shown in Figure \ref{fig:star-specs}, overlaid with the 2MASS ($J$, $H$, and $K_s$) and WISE (WISE1 and WISE2) photometry. 
The observed spectra are convolved to the 2MASS $K_s$ filter bandpass profile and flux calibrated with the $K_s$ photometry. 
For stars 2 and 4, the spectra match well with their WISE1 and WISE2 magnitudes. 
However, for stars 1 and 3, the WISE magnitudes deviate significantly from the observed spectra. 
This is possibly because the WISE photometry may be affected by saturation and confusion with nearby sources. 

\par

Determination of the continuum emission from the star, especially the photospheric features, is essential for the identification and analysis of the dust absorption features.  
All of the four target stars are late-type stars with significant CO first overtone absorption features at 2.29--2.50 \um, as well as the Ca multiplet and the Fe line in 2.261--2.267 \um.  
The observed target stars are most likely bright giant stars. 
They are bright and thus suitable for spectroscopy.  
We retrieved SpeX spectra of spectroscopic template giant stars from the IRTF Spectral Library \citep{Rayner_Infrared_2009}.  
These spectra are then reddened with a series of extinction $A_K$\footnote{Throughout this paper, we assume
$A_{K}/A_{V}=0.112$ according to \citet{Rieke_interstellar_1985}.} from 0.0 to 4.0 mag with a step of 0.01 mag following the extinction law obtained by \citet{Indebetouw_Wavelength_2005}.  
All the reddened template spectra are scaled to have the same $K_s$ magnitude as the target stars.  
Then, these scaled reddened template spectra are compared with the spectra of the target stars in 2.05--2.45 and 3.70--4.00 \um\ as well as the 2MASS $J$ and $H$ photometry to find the best-matching template spectra.  
The best-matching spectral types and the values of extinction $A_K$ are listed in Table \ref{tab:info}, and the best-matching reddened template spectra are plotted in Figure \ref{fig:star-specs}.

\par 

For star 1, the matching procedure fails, because the best-matching template spectrum shows significantly weaker absorption of the CO first overtone than that of the star 1 spectrum.  
This means that star 1 is a cooler star than the template star.  
Since the templates can hardly match the observed spectrum of star 1, we do not report the
detection of \water\ ice along this sightline.  
For stars 2 and 4, the matching result is satisfying: the CO first overtone and the 2MASS $J$
and $H$ photometry are well reproduced.  
In addition, the reddened template in the $M$-band for star 2 matches well with the observed spectra even though this wavelength range was not considered in the matching procedure.  
Significant absorption at $\sim 3$ \um, attributed to water ice, is detected towards both stars 2 and 4. 
For star 3, the template reproduces well the CO first overtone, but is slightly higher at $<2.27$ \um\ and slightly lower at $\sim 4.0$ \um\ and $>4.5$ \um\ than the observed spectrum.  
No water ice absorption is detected towards star 3, but an upper limit can be determined based on the noise on optical depth scale and the baseline uncertainty.  

\par 

By comparing with the Gaia parallaxes and the color excesses \citep{Rayner_Infrared_2009} of the template stars, we find that the distances of stars 2 and 4 are $3.9\pm 0.5$ and $3.3\pm 0.4$ kpc, respectively. 
This means that both stars are located behind SNR W44 and its associated MC, which is consistent with the water ice detection. 
We note that the spectrum of star 4 can also be fitted by a type M1+III star with a similar reduced chi-square.  
In this case, the distance of star 4 is estimated to be 1.1 kpc, which is not very likely the case considering the ice features. 
Therefore we adopt K4Ib-II as the best-fit spectral type of star 4.

\subsection{Optical depths and column densities} \label{sec:column}
\begin{figure*}[t]
    \centering
    \includegraphics[width=0.99\textwidth]{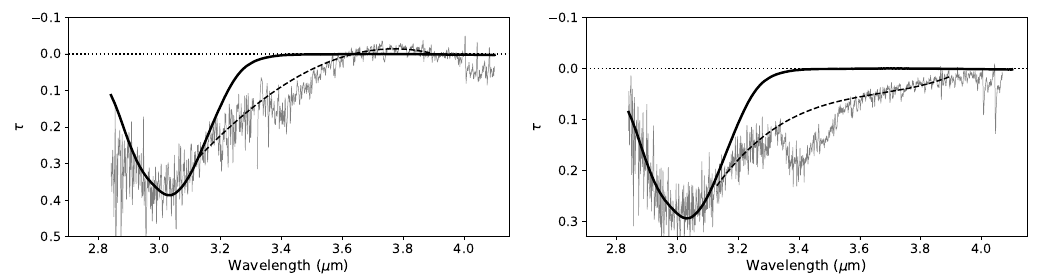}
    \caption{
    Optical depth spectra of \water\ ice towards star 2 (left panel) and star 4 (right panel) shown in grey lines. 
    The thick black lines show the best-fit \water\ ice absorption profiles obtained from \citet{Hudgins_Mid_1993}. 
    The dashed black lines are the fitted local baselines for the 3.4 \um\ absorption feature described in Section \ref{sec:column}. 
    }
    \label{fig:H2O_ice_spec}
\end{figure*}

\begin{figure*}[t]
    \centering
    \includegraphics[width=0.99\textwidth]{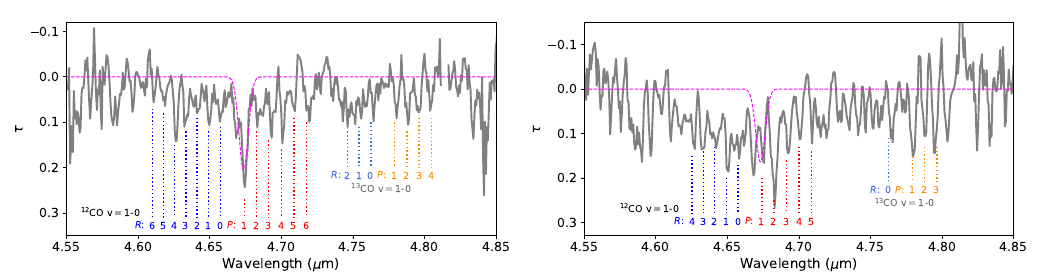}
    \caption{ $M$-band Optical depth spectra towards star 2 (left panel) and star 4 (right panel) shown in thick grey lines. 
    The dashed magenta lines show the best-fit CO ice absorption profiles using the Gaussian component of \citet{Pontoppidan_3-5_2003}. 
    CO ice is likely only detected towards star 2. 
    The possible detection of the gas phase \coa\ and \cob\ absorption lines are marked with dotted lines.
    }
    \label{fig:CO_ice_spec}
\end{figure*}

After obtaining the continuum emission, the optical depths of the absorption features can be obtained directly from $\tau=-\ln{(F_{\nu}/F_{\rm continuum})}$.  
The resulting optical depth spectra of \water\ ice and CO ice are shown in Figures \ref{fig:H2O_ice_spec} and \ref{fig:CO_ice_spec}, respectively, for stars 2 and 4.  
Note that the baseline of star 4 in the $M$-band cannot be obtained by the template fitting because the template spectrum lacks the corresponding data.  
We thus fit a local baseline to estimate the continuum for obtaining the optical depth.  
Both sightlines show significant \water\ absorption at $\sim 3$ \um. 
Gas-phase \coa\ and \cob\ absorption lines are also tentatively detected along both sightlines, among which the $P(1)$ transition of \coa\ overlaps with the CO ice absorption at 4.67 \um.    
Although we cannot differentiate between the contribution of gas- and solid-phase CO to the 4.67 \um\ absorption feature due to the limited spectral resolution, a qualitative analysis is possible. 
Towards star 2, the 4.67 \um\ absorption feature is significantly deeper than the gas-phase \coa\ absorption lines.  
Therefore, CO ice absorption is likely detected in this sightline. 
Using the expected peak wavelength of the narrow component of the CO ice band commonly observed towards dense clouds and protostars \citep{Pontoppidan_3-5_2003}, the observed CO ice absorption is redshifted by $\sim 36$ \kms.  
This is roughly consistent with the systemic velocity of W44 \citep{Yoshiike_Neutral_2013,Cosentino_Interstellar_2019} taking into account the limited velocity resolution of SpeX ($\sim300$ \kms), which suggests that the CO ice may originate from the molecular gas associated with W44.  
On the contrary, the 4.67 \um\ absorption feature towards star 4 does not show significant enhancement compared with the other gas-phase absorption lines, so it is likely the absorption feature is dominated by gas-phase absorption.

\par

The 3 $\mu$m band optical depth spectra are then fitted with the laboratory spectrum of \water\ ice at 10 K \citep{Hudgins_Mid_1993}. 
The CO ice absorptions, although confused with CO gas lines, are fitted with the empirically determined Gaussian component, with a rest wavelength at 4.6731 \um\ and an FWHM of $7.64\times10^{-3}$ \um, representing pure CO ice of \citet{Pontoppidan_3-5_2003}.  
The peak optical depths of these two features are listed in Table \ref{tab:info}.  
Because the CO ice absorption overlaps with the CO gas absorption lines, these peak optical depths should be regarded as upper limits. 
The ice column densities are estimated by
\begin{equation} \label{eq:N}
    N=\frac{1}{A}\int\tau_\nu\,d\nu
\end{equation}
where $\nu=1/\lambda$ (in $\rm cm^{-1}$) and $A$ is the integrated laboratory band strength of the absorption feature, which is $2.0\times10^{-16}\rm \ cm\ molecule^{-1}$ for the 3 \um\ \water\ ice absorption and $1.1\times10^{-17}\rm \ cm\ molecule^{-1}$ for the 4.67 \um\ CO ice absorption, both with an intrinsic uncertainty of 10\% \citep{Gerakines_infrared_1995}. 
The estimated column densities of \water\ ice and upper limits for CO ice are also listed in Table \ref{tab:info}.

\begin{figure}[t]
    \centering
    \includegraphics[width=0.48\textwidth]{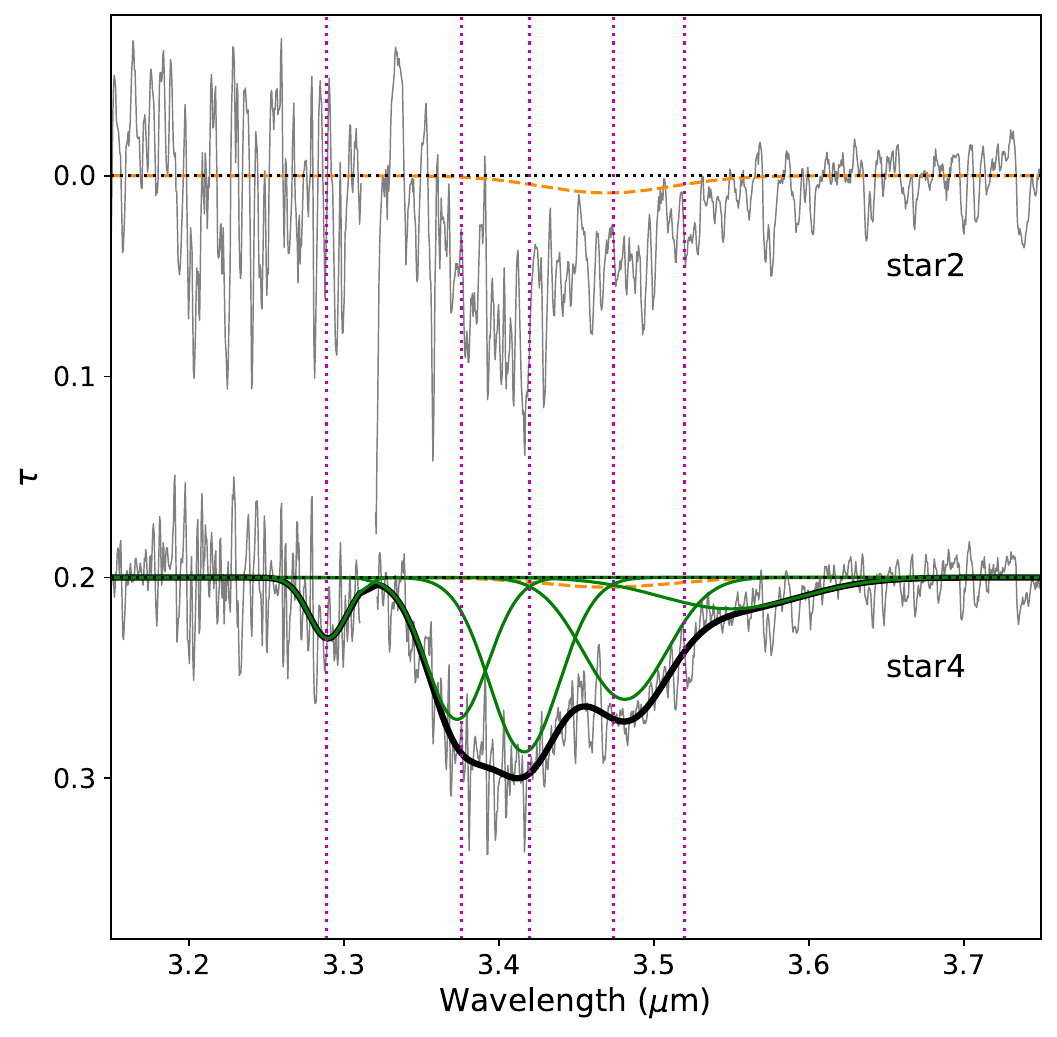}
    \caption{
    Optical depth spectra of the 3.4 \um\ absorption feature towards star 2 (upper) and star 4 (lower, offset by 0.2) shown in thin grey lines. 
    The dotted magenta lines shows the center of the $sp^3$ \chhh\ asymmetric, \chh\ asymmetric, \chhh\ symmetric, and \chh\ symmetric modes at 3.376, 3.420, 3.474, and 3.520 \um, respectively, as well as the $sp^2$ CH stretching band at 3.289 \um\ as reported by \citet{Chiar_Structure_2013}. 
    The dashed orange lines show the potential 3.47 \um\ absorption feature scaled from the water ice absorption following \citet{Brooke_New_1999}. 
    For star 4, the green lines show the results of multi-Gaussian fitting, and the thick black line shows the best-fit spectrum. 
    }
    \label{fig:3.4um_spec}
\end{figure}

From the \water\ absorption spectra, we notice excess absorption at $\sim3.4$ \um\ along both sightlines, which is believed to originate from aliphatic hydrocarbons in the interstellar dust \citep{Sandford_Interstellar_1991}.  
To study this 3.4 \um\ absorption feature in detail, we fit local baselines with a third-order polynomial at 3.13--3.25, 3.33--3.34, and 3.61--3.90 \um\ following \citet{Ishii_3.4_2002}. 
The fitted baselines are shown in Figure \ref{fig:H2O_ice_spec} in dashed black lines, and the baseline-subtracted optical depth spectra of the 3.4 \um\ absorption feature are shown in Figure \ref{fig:3.4um_spec}. 
Towards both stars, the 3.4 \um\ absorption feature is clearly detected, while it is deeper towards star 4.  
We also detect the 3.3 \um\ C-H stretch mode absorption, likely originating from aromatic molecules such as polycyclic aromatic hydrocarbon (PAH), towards star 4 \citep[e.g.][]{Chiar_Hydrocarbons_2002}.

\par

The 3.4 \um\ absorption feature of aliphatic hydrocarbons could be confused with the 3.47 \um\ feature that has been attributed possibly to ammonia hydrates \citep[e.g.,][]{Dartois_Search_2001,Dartois_Combined_2002}. 
Considering that the 3.47 \um\ absorption feature correlates very well with the \water\ absorption at 3.0 \um, we assume, according to the fitting result of \citet{Brooke_New_1999}, that the absorption profile of the 3.47 \um\ feature has a FWHM of 0.1 \um\ and a peak optical depth of
\begin{equation}
    \tau_{3.47} = (0.033\pm0.002)\,\tau_{3.0} - (0.004\pm0.004)
\end{equation}
where $\tau_{3.0}$ is the peak optical depth of \water\ ice. 
The estimated 3.47 \um\ features are displayed in Figure \ref{fig:3.4um_spec} with dashed orange lines. 

\par

To study the 3.4 \um\ feature in detail, we present multi-Gaussian fitting to decompose the optical depth spectrum towards star 4 following \citet{Chiar_Structure_2013} after the subtraction of the 3.47 \um\ feature. 
The decomposition is not conducted towards star 2 because of its relatively low S/N. 
The fitting results are shown in Figure \ref{fig:3.4um_spec} and Table \ref{tab:3.4um}. 
The four components of the 3.4 \um\ feature are attributed to the $sp^3$ \chhh\ asymmetric, \chh\ asymmetric, \chhh\ symmetric, and the \chh\ symmetric modes from low to high wavelengths. 
The results are similar to those reported towards the Quintuplet Cluster \citep{Chiar_Structure_2013} and Cygnus OB2-12 \citep{Hensley_Detection_2020}. 
One component centered at 3.29 \um\ is used to fit the PAH absorption. 
Our fitting results of the \chh\ symmetric mode at $\sim3.52$ \um\ are affected by the line wing structure at $\gtrsim 3.55$ \um, resulting in a relatively broad and red-shifted \chh\ symmetric mode compared to previous studies. 
This may cause extra uncertainty to the fitting. 
The column densities of all absorption modes are estimated according to Equation \ref{eq:N}, and the corresponding band strengths are adopted from \citet{Chiar_Structure_2013}, although the values are still under debate (see the references therein).

\begin{deluxetable}{ccccc}[t]
\tablecaption{Multi-Gaussian fitting result of the 3.4 \um\ absorption feature towards star 4
\label{tab:3.4um}}
\tablehead{
\colhead{Mode} & \colhead{\makecell{$\lambda_0$\\(\um)}} & \colhead{$\tau$} & \colhead{\makecell{$\Delta\lambda$\\(\um)}} & \colhead{\makecell{Column density\\($\rm 10^{17}\ cm^{-2}$)}} 
}
\startdata
CH ($sp^2$)           & 3.29 & 0.030 & 0.029 & 3.34 \\
\chhh\ ($sp^3$) asym. & 3.37 & 0.071 & 0.047 & 1.26 \\
\chh\ ($sp^3$) asym.  & 3.42 & 0.086 & 0.053 & 2.73 \\
\chhh\ ($sp^3$) sym.  & 3.48 & 0.060 & 0.063 & 1.40 \\
\chh\ ($sp^3$) sym.   & 3.55 & 0.016 & 0.107 & 0.95 \\
\enddata
\end{deluxetable}

\section{Discussion} \label{sec:disc}

\subsection{Line-of-sight properties towards stars 2 and 4} \label{sec:LOS}

\begin{figure}[t]
    \centering
    \includegraphics[width=0.48\textwidth]{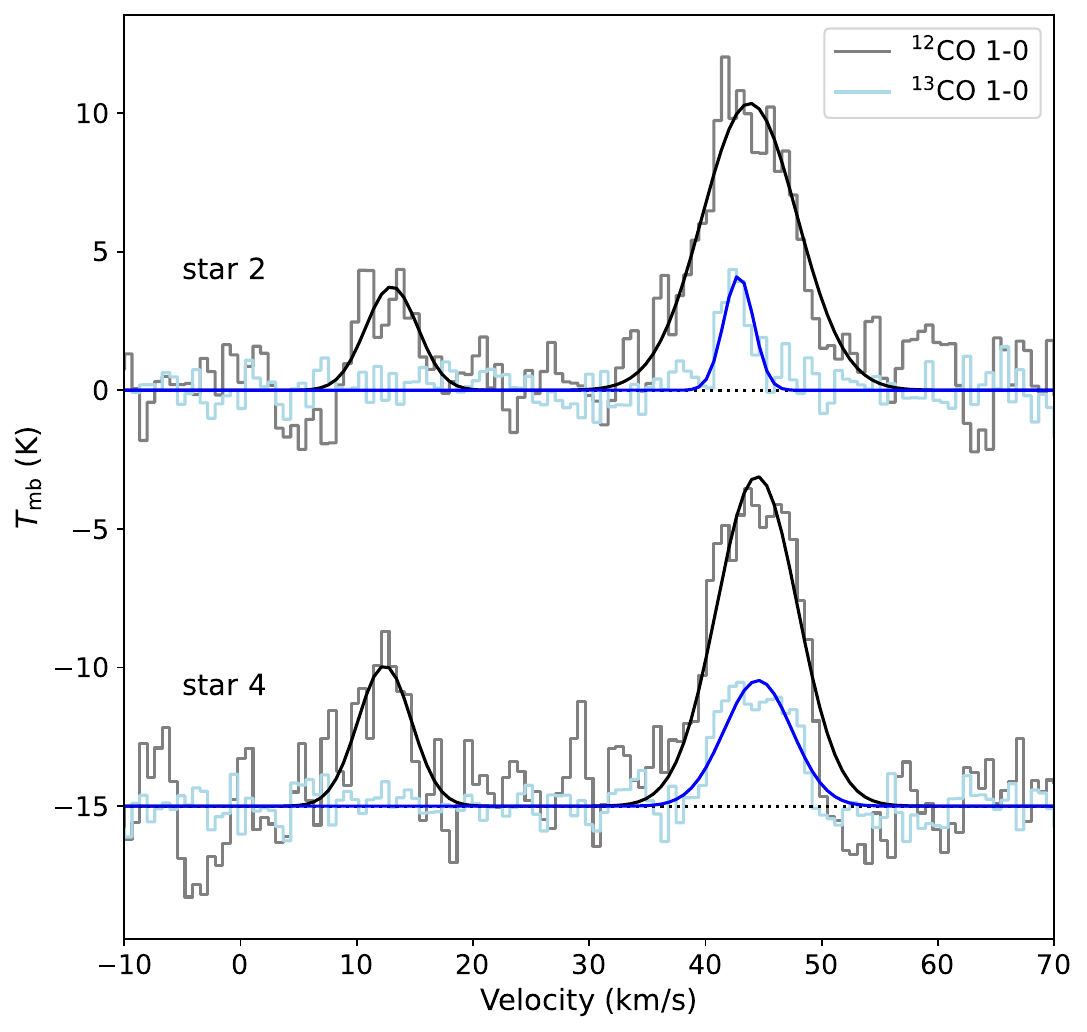}
    \caption{
    \coa\ 1--0 (grey) and \cob\ (light blue) 1--0 emission lines spectra towards stars 2 and 4 obtained from the FUGIN project \citep{Umemoto_FOREST_2017}. 
    The spectra towards star 4 are offset by 15 K. 
    The black and blue lines show the results of Gaussian fitting to the two lines, respectively. 
    }
    \label{fig:COgas}
\end{figure}
In the previous section we reported the detection of \water\ ice and the 3.4 \um\ absorption towards stars 2 and 4. 
To further study theses absorption features, we first analyze the line-of-sight properties towards these two stars. 
The distances of stars 2 and 4 are estimated to be 3.9 and 3.3 kpc, respectively (See Section \ref{sec:conti}).  
We expect both diffuse and dense clouds to contribute to the total extinction determined in Section \ref{sec:conti}. 
We need to differentiate between their contributions to interpret the observed ice and dust absorption features.

\par

To estimate the extinction which originates from the dense molecular gas, we obtain the \coa\ and \cob\ 1--0 line from the FUGIN project \citep{Umemoto_FOREST_2017} towards these two sightlines and the spectra are shown in Figure \ref{fig:COgas}. 
Both sightlines show \coa\ emission peaks at $\approx +12$ and $\approx +44$ \kms, with the latter physically associated with SNR W44 \citep{Yoshiike_Neutral_2013,Cosentino_Interstellar_2019}. 
\cob\ emission is only detected in the $\approx +44$ \kms\ component. Both components are fitted with Gaussian line profiles.  
We note that the $\approx +44$ \kms\ component of \coa\ along both sightlines shows double-peak line profiles, but the limited spectral resolution (1.3 \kms), angular resolution ($20^{\prime\prime}$ which may include two MCs in one single beam), and detection significance ($\approx 9\sigma$ in peak intensity) of the FUGIN project does not allow us to decompose them.  
This may cause an underestimation of the \cob\ column density along star 2. 
The \cob\ 1--0 line is slightly blue-shifted from the center of the \coa\ 1--0 line, suggesting that the \cob\ line is only consistent with one component of the \coa\ line. 
The other \coa\ component does not show a \cob\ counterpart and this component cannot be included in our estimation of the column density.  
We assume local thermodynamic equilibrium (LTE) and follow the method of \citet{Mangum_How_2015} to estimate the extinction caused by the $\approx +44$ \kms\ component. 
Assuming that the \coa\ 1--0 lines are optically thick, the excitation temperatures are estimated to be 13.8 and 15.3 K for the two sightlines, respectively. 
Then we find that the $A_{\rm V}$ induced by dense gas are 10.9 and 28.6, respectively, assuming $N({\rm H_2})=7\times10^5N(\rm ^{13}CO)$ \citep{Frerking_relationship_1982} and the conversion factor of \citet{Bohlin_survey_1978}. 
For the $\approx +12$ \kms\ components, we adopt the CO-to-$\rm H_2$ X factor \citep{Bolatto_CO-to-H2_2013} to estimate their \coa\ column densities and find the $A_{\rm V}$ to be 4.5 and 6.2 mag for stars 2 and 4, respectively. 
This component may be foreground or background gas not associated with W44.

\par

Diffuse dust contributing to the total extinction consists of that associated and not associated with W44. 
W44 is believed to be associated with a low-velocity H\,I bubble \citep{Yoshiike_Neutral_2013} and a high-velocity H\,I expanding shell \citep{Park_Shells_2013}. 
The column density of the low-velocity H\,I is $<2.5\times 10^{21}\rm \ cm^{-2} $ \citep{Yoshiike_Neutral_2013}, corresponding to a visual extinction of $<1.3$ mag. 
Although the high-velocity H\,I is detected in specific clumps of H\,I in W44 \citep{Park_Shells_2013}, we could not find significant ($<3\sigma$) emission of high-velocity H\,I from the GALFA-H\,I data towards stars 2 and 4. 
We estimate the column density of the high-velocity H\,I to be $2n_0R_{\rm s}\approx1.5\times10^{20}\rm \ cm^{-2}$ where $n_0\approx1.9 \rm \ cm^{-3}$ is the H\,I density and $R_{\rm s}\approx12.5$ pc is the radius of the high-velocity shell obtained by \citet{Park_Shells_2013}. 
This column density corresponds to a visual extinction of $\sim 0.08$ mag.  So the total extinction from the diffuse dust associated with W44 is $<1.4$ mag. 
To assess extinction caused by the diffuse dust not associated with W44, we refer to the average visual extinction in Galactic diffuse clouds, 1.8 mag ${\rm kpc}^{-1}$ \citep{Whittet_Dust_1992}, which corresponds to 7.0 and 5.9 towards stars 2 and 4 respectively.  
However, there might be local fluctuation around this average value. 
The latest version of the Bayestar dust map based on Pan-STARRS 1, 2MASS and Gaia \citep{Green_3D_2019}, which is sensitive to extended diffuse dust, gives $A_{V}\sim 4.9$ and 4.2 mag, respectively.  
We adopt these values for the following analysis.  
Another three-dimensional (3D) dust map, based on Gaia and LAMOST \citep{Wang_All-sky_2025}, predicts $A_{V}\sim 4.0$ and 3.6 mag, respectively, which is roughly consistent with the previous one.

\par

From the analysis above, we derive the visual extinction towards stars 2 and 4 to be $\sim 17.2$ and $\sim 34.2$, respectively. 
Note that the extinction caused by the molecular gas at $\sim +44$ \kms\ towards star 2 is likely underestimated. 
The derived extinction is lower than that obtained from the spectral template star matching (22.9 mag, see Table \ref{tab:info}) for star 2, and higher for star 4 (23.1 mag). 
Provided the uncertainties in our analysis, we consider that this analysis is indicative of the contribution of different phases of dust clouds in the lines of sight. 
We can draw a qualitative conclusion that dense molecular gas at $\approx +44$ \kms\ contributes a significant portion of the total extinction for both stars 2 ($\sim 63\%$) and 4 ($\sim 84\%$). 
The diffuse dust along the line of sight, not associated with W44, contributes $\sim28\%$ and $\sim 12\%$ to the total extinction for stars 2 and 4, respectively, while the contribution from the H\,I cloud associated with W44 shell is almost negligible.  
But we note that our analysis is coarse because we ignore the fact that the near-infrared absorption observations, archival millimeter CO 1--0 line data, and the 3D dust extinction maps trace interstellar gas at different angular scales.  
So there might be small-scale structures that are not probed by the CO lines and the extinction maps. 
And possibly, though unlikely given the derived distances, the stars may be located inside the $\approx +44$ \kms\ MC, so not the entire MC contributes to the total dense cloud extinction derived from the millimeter-wave CO 1-0 observations.

\subsection{The \water\ and CO ice} \label{sec:disc_ice}

\begin{figure*}[t]
    \centering
    \includegraphics[width=0.99\textwidth]{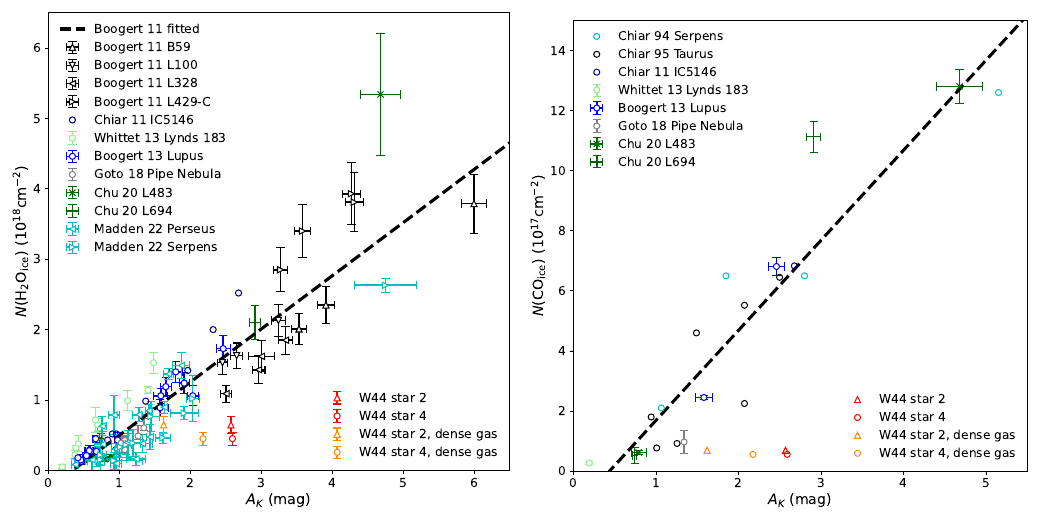}
    \caption{ Correlation plots of $A_{K}$ with column densities of \water\ (left panel) and CO (right panel) ices.  
    The data obtained towards stars 2 and 4 are shown in red triangles and circles, respectively (note that the obtained column densities of the CO ice should be regarded as upper limits; Section \ref{sec:column}), while the orange triangles and circles are those where the $A_{K}$ from diffuse gas is subtracted.  
    The other data points are taken from \citet{Chiar_Solid_1994,Chiar_High-Resolution_1995,Boogert_Ice_2011,Chiar_Ices_2011,Whittet_Ice_2013,Boogert_Infrared_2013,Goto_first_2018,Chu_Observations_2020,Madden_Infrared_2022} and are marked in the label of each panel.  
    The dashed black line in the left panel shows the best-fit $A_{K}$-$N({\rm H_2O})$ linear relation in \citet{Boogert_Ice_2011}, while in the right panel shows the best-fit $A_{K}$-$N({\rm CO})$ linear relation from all archival data points in the figure.  }
    \label{fig:ice_AK}
\end{figure*}

According to Section \ref{sec:column}, \water\ ice is detected towards both stars 2 and 4. 
CO ice is also possibly detected, but due to its overlap with the CO gas absorption lines, the estimated column densities of CO ice should be regarded as upper limits. 
In Figure \ref{fig:ice_AK}, we show the $N({\rm H_2O_{ice}})$-$A_{K}$ and $N({\rm CO_{ice}})$-$A_{K}$ correlation in our study and previous observations in nearby MCs. 
Previous studies have revealed a clear linear correlation between the ice column densities and extinctions \citep[e.g.,][]{Boogert_Ice_2011}.  
However, as shown in Figure \ref{fig:ice_AK}, the results of our observations are well below the best-fit line in nearby MCs. 
This has also been found towards the Galactic Center \citep{Jang_Massive_2022}. 
One of the possible explanations could be that both diffuse and dense gas contribute to the total extinction along the sightlines, but only dense molecular gas can harbor the \water\ and CO ice. 
At low extinctions and in low-density regions, the ices are photodissociated and photodesorbed \citep[e.g.,][]{Whittet_Interstellar_1983}. 
However, the ISM towards stars 2 and 4 consists mainly of dense molecular gas according to our analysis in Section \ref{sec:LOS}. 
We also show in Figure \ref{fig:ice_AK} the data points towards stars 2 and 4 with the extinction from diffuse gas is subtracted, and so only extinction from dense gas is considered. 
The \water\ ice column density towards star 2 is roughly comparable to previous observations (lower by a factor of 1.5), while towards star 4, the \water\ ice column density is still lower by a factor of 3 compared to nearby MCs.  
Therefore, simply correcting for the diffuse cloud extinction cannot explain the low ice abundances, especially for star 4.

\par

Perhaps the ices are destroyed by the shocks of SNR W44.  
Destruction of dust in MCs interacting with the shock of W44 has been revealed by SiO gas emission \citep{Cosentino_Widespread_2018,Cosentino_Interstellar_2019}, and simulations have found that shocks in MCs can release the \water\ and CO ice back to the gas phase \citep[e.g.,][]{Burkhardt_Modeling_2019}. 
We also note that the abundance of CO ice relative to \water\ ice is $\lesssim12\%$ along both sightlines. 
The range of this value observed in quiescent molecular clouds and cores is 9\%--67\% (see Table 2 of \citet{Boogert_Observations_2015}).  
Thus, the observed ice abundance towards W44 are rather low compared to quiescent clouds. 
A natural explanation for the observed low CO ice abundance is that after the shock sublimates the ices, the \water\ ice freezes out earlier than CO ice in the cooling phase of the shock. 
CO is much more volatile than \water. 
Higher spectral resolution observations with higher S/N have shown that the CO ice at $\sim 4.67$ \um\ consists of three components, each with different sublimation temperatures: pure CO ice, CO in polar species, and CO in apolar species, and the ratio among them can be a tracer of the ice temperature \citep[e.g.,][]{Tielens_Physics_1994}.  
Further observations of these features toward W44 can help identifying whether the shock passage has affected the ice temperature and composition.

\par

We note, from Figure \ref{fig:13COmap}, that star 4 is projected $\sim 20^{\prime\prime}$ ($\sim0.3$ pc at a distance of 3 kpc) outside the radio boundary of W44. 
However, the magnetic precursor of the SNR shock in MCs can affect the molecular gas upstream of the radio emission \citep{Draine_Magnetohydrodynamic_1983,Draine_Theory_1993}, and can also release the icy species to the gas phase \citep{Jimenez-Serra_Parametrization_2008}. 
The length scale of the magnetic precursor can be estimated from \citep{Draine_Physics_2011,Zhou_Unusually_2022b}: 
\begin{equation}
    l_{\rm mag}\sim 3.4\times 10^{19} b_0^2 \left[ \left( \frac{n_{\rm H}}{\rm cm^{-3}} \right) \left( \frac{V_{\rm sh}}{\rm km\, s^{-1}} \right)^2 \right] ^{-0.5} \ \rm cm, 
\end{equation}
where $b_0 = (B/{\rm \mu G}) / (n_{\rm H}/{\rm cm^{-3}})^{0.5}$ and $V_{\rm sh}$ is the shock velocity.  
For the shock in MCs associated W44, typical values are $n_{\rm H}\sim 10^4\rm \ cm^{-3}$, $V_{\rm sh}\sim 23$ \kms, and $B\sim 880\rm \ \mu G$ \citep{Cosentino_Interstellar_2019}. 
We find that the typical $l_{\rm mag}$ is $\sim 0.4$ pc. 
This is roughly consistent with the simulation results of \citet{Cosentino_Interstellar_2019}, and of similar order as the projected distance of star 4 from the radio boundary of W44 ($\sim 0.3$ pc).  
Therefore, although star 4 is projected outside the radio boundary of W44, the MC along this sightline can still be affected by the magnetic precursor of the SNR shock.

\par

In addition, the CR-induced desorption can also play a role in the low CO ice abundance. 
According to the estimation of \citet{Tu_Yebes_2024}, the timescale of CR-induced non-thermal desorption of CO ice is $\sim 3\times 10^4$ yr assuming that the CR ionization rate is $\rm \sim 10^{-15}\ s^{-1}$, the best-fit value obtained by \citet{Cosentino_Interstellar_2019}. 
This timescale is comparable to the age of W44 (see Section \ref{sec:intro}). 
Since the desorption rate is exponentially correlated to the desorption barrier \citep{Hasegawa_New_1993}, and the desorption barrier of \water\ ice higher than that of CO ice by a factor of 3--4 \citep{Wakelam_Binding_2017}, CR-induced non-thermal desorption is negligible to \water\ ice on the timescale of the SNR. 
This could also lead to the low CO ice abundance we observe.

\subsection{Origin of the aliphatic hydrocarbon absorption} \label{sec:disc_dust} 

\begin{figure}[t]
    \centering
    \includegraphics[width=0.48\textwidth]{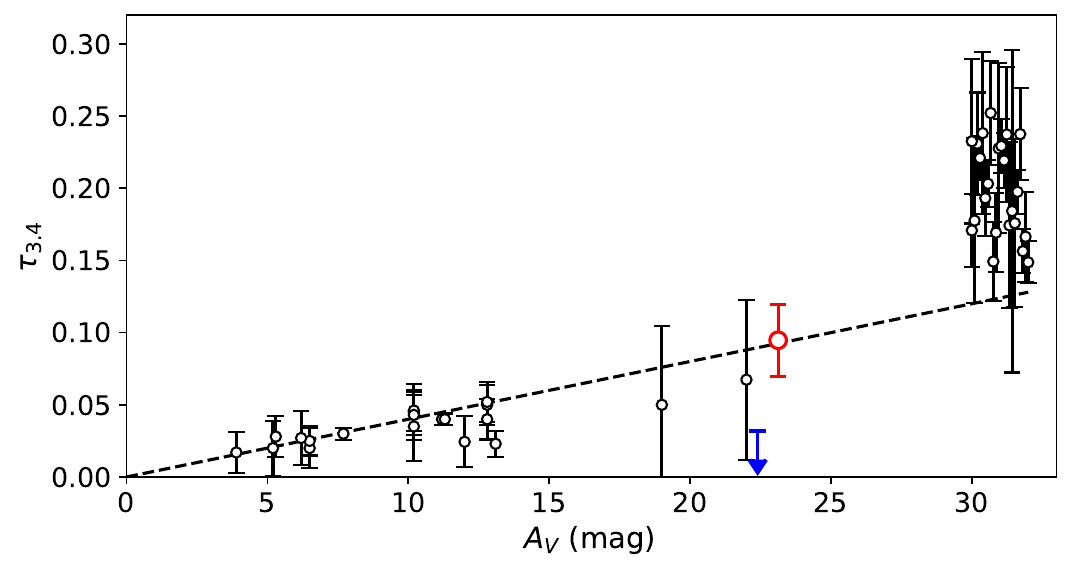}
    \caption{ Optical depth of the aliphatic hydrocarbon absorption at 3.4 \um\ ($\tau_{3.4}$) against the visual extinction ($A_{V}$). 
    The black data points are taken from previous observations \citep{McFadzean_Infrared_1989,Adamson_3.4-mu.m_1990,Sandford_Interstellar_1991,Pendleton_Near-Infrared_1994,Chiar_Composition_2000,Chiar_Circumstellar_2001,Chiar_Hydrocarbons_2002,Rawlings_Infrared_2003,Dartois_Diffuse_2004} which were in turn compiled by \citet{Godard_deep_2012}. 
    The data points at around $A_{V}\sim 30$ mag are from Galactic Center, while the others are from local diffuse clouds.  The blue arrow shows the upper limit obtained in the Taurus MC along the sight-line of Elias 16 \citep{Chiar_Three_1996}.  
    The observed value towards star 4 is shown in open red dot. 
    The dashed black line shows $A_{V}/\tau_{3.4}=250$ as obtained by \citet{Pendleton_Near-Infrared_1994} in local diffuse clouds.  }
    \label{fig:3.4_AV}
\end{figure}

The 3.4 \um\ absorption feature has been attributed to the absorption from aliphatic hydrocarbon in diffuse interstellar clouds \citep{Sandford_Interstellar_1991}.  
Previous studies found, as shown in Figure \ref{fig:3.4_AV} where we plot the optical depths of the 3.4 \um\ absorption feature $\tau_{3.4}$ against the visual extinctions $A_{V}$, that the $A_{V}/\tau_{3.4}$ ratios are different between local diffuse clouds ($\sim 250$) and the sightlines towards the Galactic Center ($\sim 150$) \citep[e.g.,][]{Pendleton_Near-Infrared_1994,Sandford_Galactic_1995}. 
Recent observations found even lower values ($\sim 80$) in specific regions towards the Galactic Center \citep{Gunay_Mapping_2022}.
Most of the detections of the aliphatic hydrocarbon absorption were associated with the diffuse clouds, while in most dense MCs, only upper limits are given \citep[e.g.,][]{Chiar_Three_1996}.  
Some of these upper limits are well below the diffuse medium trend. 
A recent study of \citet{Jang_Massive_2022} proposed that molecular gas in the Galactic Center may have a non-negligible contribution to the total $\tau_{3.4}$ because it shows a good correlation with the \water\ ice absorption.

We also plot our result towards star 4 in Figure \ref{fig:3.4_AV}.  
It is almost consistent with previous results showing $A_{V}/\tau_{3.4}\approx250$ in local diffuse clouds.  
However, the total extinction of $A_{V}\sim 23$ towards star 4 is dominated by dense molecular clouds associated with W44 (Section \ref{sec:LOS}). 
If the 3.4 \um\ feature originates from the diffuse dust along the line of sight unrelated to W44, its absorption strength is $\sim 4$ times larger than what is typically found in diffuse dust clouds. 
Here we propose several scenarios to qualitatively explain the enhanced 3.4 \um\ absorption towards star 4. 

\par

1. The elevated absorption of 3.4 \um\ could be due to an enhanced abundance of aliphatic hydrocarbon in the diffuse dust associated with SNR W44. 
According to \citet{Chiar_Structure_2013}, shock processing in diffuse ISM is an important pathway of the formation of aliphatic hydrocarbon dust, via impinging H atoms to amorphous carbon dust grains \citep{Tielens_Physics_1994}. 
The required shock velocity \citep[$\sim 100$ \kms,][]{Chiar_Structure_2013} can be reached by W44, which has driven an fast expanding H\,I shell at a velocity of $\sim 135$ \kms\ \citep{Park_Shells_2013}, even if it is a middle-aged SNR.  However, according to our analysis in Section \ref{sec:LOS}, dust in the high-velocity H\,I shell contributes only $\sim 0.08$ mag of extinction. 
This means that the abundance of aliphatic hydrocarbon in this shell should be enhanced by a factor of $> 200$ to account for the observed $\tau_{3.4}$.

\par

2. The enhanced absorption of 3.4 \um\ could be due to an elevated abundance of aliphatic hydrocarbon in the molecular cloud associated with the W44 SNR.  
Although aliphatic hydrocarbon dust has seldom been detected in MCs, the reason for this is still unknown.  
According to \citet{Godard_Ion_2011}, the formation of aliphatic hydrocarbon is hampered in molecular gas instead of diffuse gas due to (a) low flux of H atoms, (b) low temperature which decelerates the chemical reaction, and (c) the formation of ice mantle which stops the impinging H atoms from hitting the carbon dust cores. 
However, these effects of molecular gas can be alleviated by the shock and CRs.  
For case (a), CR ionization can transfer \hh\ to atomic H in MCs \citep{Padovani_Production_2018}, which can in turn be accelerated by the shock to reach high flux of H atoms.  
For case (b), observations have shown that the shock of SNR W44 can heat the molecular gas to a temperature of $\approx 60$ K \citep{Anderl_APEX_2014}, and higher temperatures have been observed in shocked MCs towards other SNRs \citep[e.g., $\sim 200$ K in IC443,][]{vanDishoeck_Submillimeter_1993}.  
This temperature is high enough to induce a high formation rate of aliphatic hydrocarbons \citep{Mennella_Activation_2006}.  
For case (c), the ice mantles can be partially destroyed by the shock as we have shown in Section \ref{sec:disc_ice}, allowing the H atoms to hit the carbon dust cores. 
Therefore, it is possible that the shocks and CRs may induce a suitable physical environment for the formation of aliphatic hydrocarbon dust.

\par

Another possible route to produce aliphatic hydrocarbon in MCs could be the CR processing of ice mantles.  
Although CRs are the main source of aliphatic hydrocarbon destruction in MCs, experiments have found that the destruction rate of aliphatic hydrocarbon by CRs is not high enough to explain the absence of 3.4 \um\ absorption in MCs \citep[e.g.,][]{Godard_Ion_2011,Mate_High-energy_2016}. 
On the other hand, the processing of ice composed of small hydrocarbons like $\rm CH_4$ by UV photons can lead to the formation of refractory hydrocarbon dust \citep[e.g.,][]{Dartois_Diffuse_2004,Lo_Formation_2021}. 
Considering the similarity between CRs and UV photons in ice processing \citep{Arumainayagam_Extraterrestrial_2019} and that CRs can induce UV photons in MCs \citep[e.g.,][]{Prasad_UV_1983,Sternberg_Cosmic-Ray_1987}, we propose that CRs may induce the formation of aliphatic hydrocarbon through the processing of hydrocarbon ice mantle in MCs.

\par

3. The high value of $\tau_{3.4}$ in Figure \ref{fig:3.4_AV} could be due to an underestimation of the visual extinction $A_{V}$. 
Shock processing can significantly change the structure and composition of the dust \citep[e.g.,][]{SerraDiaz-Cano_Carbonaceous_2008}, affecting the extinction curve, and in turn misleading our estimation of the extinction. 
\citet{Gao_understanding_2010} suggested that the low $A_{V}/\tau_{3.4}$ towards the Galactic Center could be due to the enhanced porosity of dust grains. 
Since the ISM shocked by SNRs have similar physical conditions to the ISM in the Galactic Center (both are shocked with enhanced CR ionization rates \citep{Oka_Central_2019}), it is also possible that the extinction curve has been altered towards star 4.

\par

All these three scenarios could qualitatively explain the enhanced 3.4 \um\ absorption towards star 4.  However, these are only preliminary discussions. 
Further analysis of the line-of-sight composition towards star 4, the extinction, and the experimental properties of aliphatic hydrocarbons are needed to draw a final conclusion.

\section{Conclusion} \label{sec:con}
We performed $K$, $L$, and $M$-band spectroscopy with IRTF/SpeX towards four stars expected to be absorbed by dust and ices in the giant MC interacting with SNR W44.  
Our main findings are:

\begin{enumerate}

\item The continuum level for the dust and ice absorption features was
  determined by matching the observed spectra with reddened template
  spectra. The spectral types and extinctions were successfully
  estimated for stars 2, 4 (both $A_K\sim 2.6$; $A_V\sim 23$), and 3
  ($A_K\sim 1.6$; $A_V\sim 14$) while the templates could hardly be
  matched to the spectrum of star 1.

\item \water\ ice absorption around 3.0 \um\ and aliphatic hydrocarbon dust absorption around 3.4 \um\ was detected towards both stars 2 and 4.  
We also detected the PAH absorption at 3.29 \um\ towards star 4.  
CO ice absorption at 4.67 \um\ was likely detected towards star 2, but was contaminated by CO gas absorption lines at similar wavelengths.

\item After separating the contributions of the extinction by diffuse dust along the lines of sight and the extinction local to the W44 molecular cloud, we find that the \water\ ice column densities are lower by a factor of 1.5--3 compared to nearby MCs at similar dense cloud extinctions.  
This is probably because of the destruction of the ice mantles by shock and CRs from SNR W44.  
The $N({\rm CO_{ice}})/N({\rm H_2O_{ice}})$ ratio was estimated to be $\lesssim 12\%$ towards both sightlines, which further hints the destruction by shock and CRs because CO ice could be destroyed more easily by shocks while \water\ ice could freeze out earlier in the cooling phase of the shock.

\item The 3.4 \um\ absorption feature from aliphatic hydrocarbon dust is remarkably prominent towards star 4, even more so considering that the extinction in this sightline mainly comes from the dense molecular cloud associated with W44. 
If this feature originates in the unrelated diffuse dust along the line of sight, it is a factor of $\sim 4$ stronger than expected.  
If the 3.4 \um\ absorption feature originates in the diffuse dust shell associated with W44, then its depth would be a factor of $\gtrsim 200$ deeper than expected from the general $A_V$-$\tau_{3.4}$ dust relation in the galactic disk.  
If it is, alternatively, from the dense dust, then the depth of the absorption is consistent with this relation, although in previous work the 3.4 $\mu$m absorption feature was not found to be associated with dense cloud material.  
We discuss several possibilities for this enhancement, including: (1) the enrichment of aliphatic hydrocarbons by SNR shocks in the diffuse cloud associated with W44; (2) the formation of aliphatic hydrocarbons in the molecular cloud due to the suitable physical environment produced by shock and CRs, or CR processing of ices in the dust mantle; (3) a change in extinction law due to SNR interaction which could mislead our estimation of the extinction.

\end{enumerate}  

\begin{acknowledgments}

The authors thank the staff crew of the IRTF for their assistance in obtaining the data. 
T.-Y. Tu thanks Marie Godard, Xuejuan Yang, Ping Zhou and Wenjin Yang for helpful discussion. 
This work is supported by National SKA Program of China (2025SKA0140100) and NSFC grants Nos. 12121003, 12573047, and 12173018. 
The observation was carried out by T.-Y. Tu remotely with the Infrared Telescope Facility, which is operated by the University of Hawaii under contract 80HQTR24DA010 with the National Aeronautics and Space Administration.

\end{acknowledgments}

\vspace{5mm}

\facilities{IRTF, No: 45m, MeerKAT, Arecibo}

\software{Astropy \citep{AstropyCollaboration_Astropy_2018, AstropyCollaboration_Astropy_2022}, 
          Matplotlib (\url{https://matplotlib.org})), 
          Spextool \citep{Cushing_Spextool_2004}
          }

\bibliography{2025_article_W44_ice}{}
\bibliographystyle{aasjournal}

\end{CJK*}
\end{document}